\documentclass[12pt]{article}
\pdfoutput=1

\usepackage{amsmath}
\usepackage{amsfonts}
\usepackage{amssymb}
\usepackage{float}
\usepackage{ulem}
\usepackage{multirow}

\usepackage[latin1]{inputenc}

\usepackage{comment}

\usepackage{hyperref}
\newlength{\abstractwidth}
\usepackage{color}
\usepackage{graphicx}

\renewcommand{\thefootnote}{\fnsymbol{footnote}}
\renewcommand{\thanks}[1]{\footnote{#1}}
\newcommand{\starttext}{
\setcounter{footnote}{0}
\renewcommand{\thefootnote}{\arabic{footnote}}}

\newcommand{\bea}{\begin{eqnarray}}
\newcommand{\eea}{\end{eqnarray}}
\newcommand{\ee}{\end{equation}}
\newcommand{\be}{\begin{equation}}
\newcommand{\<}{\langle}
\renewcommand{\>}{\rangle}

\newcommand{\no}{\nonumber}

\DeclareMathOperator{\sign}{sign}

\def\cN{{\cal N}}
\def\cO{{\cal O}}

\def\a{\alpha}

\def\k{\kappa}

\def\n{\nu}
\def\m{\mu}
\def\o{\omega}

\def\D{\Delta}
\def\G{\Gamma}

\def\no{\nonumber}

\long\def\symbolfootnote[#1]#2{\begingroup%
\def\thefootnote{\fnsymbol{footnote}}\footnote[#1]{#2}\endgroup}

\begin{document}

\begin{titlepage}

\begin{center}

\vskip 2cm
{\Large \bf Holographic two-point functions for Janus interfaces in the $D1/D5$ CFT.}

\vskip 1.25 cm {\bf Marco Chiodaroli,${}^\blacklozenge$ John Estes,${}^\clubsuit$ and Yegor Korovin${}^\bigstar$}\\

{\vskip 0.5cm \textit{ \small
${}^\blacklozenge$Department of Physics and Astronomy, \\
Uppsala University, \\
SE-75108 Uppsala,
Sweden}\\
marco.chiodaroli@physics.uu.se}\\

{\vskip 0.5cm \textit{ \small
${}^\clubsuit$Department of Physics,\\
Long Island University,\\
1 University Plaza, Brooklyn, NY 11201, United States}\\
John.Estes@liu.edu
}\\

{\vskip 0.5cm \textit{ \small
${}^\bigstar$Max-Planck-Institut f{\"u}r Gravitationsphysik,\\
Albert-Einstein-Institut, \\
Am M{\"u}hlenberg 1, 14476 Golm, Germany}\\
jegor.korovins@aei.mpg.de }

\end{center}

\vskip 1 cm

\begin{abstract}
\baselineskip=16pt
This paper investigates  scalar perturbations in the top-down supersymmetric Janus solutions
dual to conformal interfaces in the $D1/D5$ CFT, finding analytic closed-form solutions.
We obtain an explicit representation of the bulk-to-bulk propagator and extract the two-point correlation function
of the dual operator with itself, whose form is  not fixed by symmetry alone.
We give an expression involving the sum of conformal blocks associated
with the bulk-defect operator product expansion and briefly discuss finite-temperature extensions.
To our knowledge, this is the first two-point function computation
for a fully-backreacted, top-down holographic defect.
\end{abstract}

\end{titlepage}

\setcounter{tocdepth}{2}

\tableofcontents

\starttext
\setcounter{footnote}{0}


\section{Introduction}

The structure of conformal field theories (CFTs) can be enriched by introducing  boundary conditions which break part of the conformal symmetry.
A simple way to do so is to add a codimension-one planar boundary or interface which preserves
a $SO(d,1)$ subgroup of the $SO(d+1,1)$ conformal group.
Such extended objects are ubiquitous in string theory and condensed matter physics and are referred to as conformal defects.

In the presence of defects, conformal symmetry poses less stringent constraints on the structure of the correlation functions.
The two-point  function is no
longer completely fixed by symmetry.
Rather, it is an arbitrary function of an invariant cross-ratio. The implications of conformal invariance
in the presence of defects have been studied extensively \cite{Cardy:1984bb, McAvity:1995zd, Billo:2016cpy, Gadde:2016fbj},
most recently in the context of quantum information theory, where
a proof of the g-theorem for boundary CFTs in $1+1$ dimensions has been obtained
together with its higher-dimensional analogs
 \cite{Friedan:2003yc, Jensen:2015swa, Casini:2016fgb}.

In the context of the  $AdS_{d+1}/$CFT${}_d$ correspondence,
the simplest holographic realizations of defects are obtained
by adding $AdS_d$ probe branes into the $AdS_{d+1}$ bulk geometry \cite{Bachas:2000fr,Bachas:2001vj}
or by introducing an additional boundary in the bulk supergravity solution \cite{Takayanagi:2011zk,Fujita:2011fp,Erdmenger:2014xya}.
In such setups, one can successfully model many features of condensed matter systems
such as the Kondo model \cite{Erdmenger:2013dpa,OBannon:2015cqy, Erdmenger:2015spo, Erdmenger:2016vud,Erdmenger:2016jjg}.

Obtaining backreacted analytic solutions dual to defects is a considerably more involved enterprise.
A non-supersymmetric ``Janus" solution dual to an interface in a four-dimensional bulk CFT was first constructed in ref. \cite{Bak:2003jk}.
This solution has a geometry obtained  by foliating the spacetime
with $AdS_4$ slices and gives a non-singular dilatonic deformation of
the $AdS_5 \times S^5$ vacuum of type IIB supergravity in which the dilaton smoothly interpolates between two different asymptotic
values (the two ``faces" of Janus). It is interpreted as the holographic dual of two four-dimensional
CFTs with different coupling constants which are glued together along a codimension-one interface \cite{Clark:2004sb}.

Over the years, the Janus solution has produced a variegated offspring.
Supersymmetric extensions that possess asymptotic regions where the geometry is
locally $AdS_5 \times S^5$  were first obtained in refs.
\cite{D'Hoker:2006uu,D'Hoker:2007xy, D'Hoker:2007xz} and further studied in \cite{Aharony:2011yc,  Benichou:2011aa, Clark:2013mfa}.
The corresponding solutions in $\cN = 2$ gauged supergravity in five dimensions
and their  embedding in $\cN = 8$, five-dimensional gauged supergravity were discussed in
\cite{Clark:2005te,Suh:2011xc}.
Gaiotto and Witten subsequently classified the intersecting-brane configurations corresponding to
the supersymmetric Janus solutions and extended the solutions to nonzero theta angle \cite{Gaiotto:2008ak,Gaiotto:2008sd}.
Later work also uncovered Janus solutions in M-theory \cite{D'Hoker:2008wc,D'Hoker:2008qm,Berdichevsky:2013ija}
and attained a  classification relying on superconformal algebras
\cite{D'Hoker:2008ix}.
Non-supersymmetric solutions with asymptotic locally-$AdS_3 \times S^3$
geometries were constructed in ref. \cite{Azeyanagi:2007qj}.
The supersymmetric $AdS_3 \times S^3$ Janus solutions
which we will utilize in this paper were first obtained in ref. \cite{Chiodaroli:2009yw}
and further studied in refs. \cite{Chiodaroli:2010mv,Chiodaroli:2009xh,Chiodaroli:2010ur} together with large
classes of string-junction solutions which possess
more than two asymptotic regions.
The solutions obtained in \cite{Chiodaroli:2009yw}
have geometries given by a $AdS_2\times S^2 \times M_4$ fibration over a two-dimensional
base space, where $M_4$ is either the four-torus or a $K3$ manifold,
and, in the simplest cases, are holographic duals to interfaces in the $D1/D5$ CFT.
Analogous solutions  in Roman's type  4b six-dimensional
supergravity were constructed in refs. \cite{Chiodaroli:2011nr,Chiodaroli:2011fn,Chiodaroli:2012vc}.
Finally, the literature has discussed various finite-temperature versions
of Janus solutions either numerically \cite{Bak:2011ga} or analytically \cite{Estes:2015jha}.

The advantage of top-down constructions is that often the dual (deformed) field theory is known explicitly.
The drawback is that these supergravity solutions are considerably more complicated than their artificial bottom-up relatives.
In particular,  most holographic computations conducted  to date
with top-down solutions have focused on boundary or entanglement entropy \cite{Azeyanagi:2007qj,Chiodaroli:2010mv,Bak:2013uaa,Estes:2014hka,Gutperle:2015hcv,Gutperle:2016gfe},
for which there exist a simple holographic prescription due to Ryu and Takayanagi \cite{Ryu:2006bv} (see also \cite{Jensen:2013lxa, Korovin:2013gha, Bianchi:2016xvf} for non-supersymmetric examples).
It
seems particularly difficult to extract any correlation
function whose form is not completely fixed by the residual defect conformal symmetry.

In this paper, we take a concrete step towards the computation of more general holographic observables in Janus backgrounds, and
study one particular family of top-down defects for which we are able to perform explicitly a number of analytic computations.
We identify a particular field which decouples from other fields, solve the linearized
field equations, construct the bulk-to-bulk propagator and obtain the two-point correlator,
which a priori is an arbitrary function of the conformal crossratio.
To our knowledge, this is the first computation of this kind for a fully-backreacted, top-down holographic defect.

One interesting feature of our holographic calculation is that the final result for the two-point function is explicitly
expressed in terms of the conformal blocks associated with the boundary operator product expansion (BOPE).
As we review in the next section, there are two kinds of operator product expansions (OPEs) in the presence of a defect.
One is the usual OPE, which can be used when two operators are close to each other but stay away from the defect.
The second kind of OPE, the BOPE, is relevant when an operator is brought close to the defect but is separated from the other insertions.
It is the latter BOPE which is made explicit in our holographic computation. Of course, writing a two-point function as a sum of conformal
blocks is not automatically consistent.  As a non-trivial check of our final result, we have  partially resummed the
BOPE conformal blocks and reproduced the identity contribution in the crossed channel, i.e. the identity contribution to the ordinary OPE.

This paper is organized as follows. In the next section, we collect some relevant facts about defect CFTs.
In section \ref{solution}, we discuss the background supergravity solution and scalar perturbations.
In section \ref{linearmodes}, we solve the linearized field equations for the decoupled scalar field.
Finally, in section \ref{propcorr}, we construct the bulk-to-bulk propagator from the linear modes and compute the two-point correlation
function of the dual operator.

\section{Basic facts about defect CFTs}

Here we review the basic facts about codimension-one defects.
For a more detailed field-theory discussion we refer to \cite{McAvity:1995zd, Billo:2016cpy}.

For definiteness, we consider an Euclidean CFT on $\mathbb{R}^d$. A conformal defect of
codimension one preserves $SO(d,1)$ defect conformal symmetry. The coordinates along the defect are labelled as $x^a$ with $a=1,\ldots,d-1$,
whereas those in the bulk are labelled as $x^{\m} = (x^a, x_{\perp})$.

In the presence of a defect, one distinguishes between bulk and defect excitations. The defect does not affect the fusion of
primary operators away from the defect  (the so-called bulk operators). Thus we have the usual bulk OPE channel
\be
\cO(x_1) \cO(x_2) = \sum_k \frac{c_{12k}}{(x^2_{12})^{(\D_1 + \D_2 - \D_k + J)/2}} C^{(J)}(x_{12}, \partial_2)_{\m_1 \ldots \m_J} \cO_k^{\m_1 \ldots \m_J} (x_2),
\ee
where $C^{(J)}$ are some known differential operators.

Any bulk operator $\mathcal{O}$ can be restricted to a defect. In general, the defect possesses its
own local excitations as well. We will refer to both kinds of operators as the defect operators $\hat{\cO_i}$. When a
bulk operator is brought close to the defect, it becomes indistinguishable from a defect operator.
This is captured by the bulk-to-defect OPE (also referred to as BOPE),
\be
\label{qftbope}
\cO(x) = \sum_{\hat{\cO}} \frac{b_{\cO \hat{\cO}}}{|x_{\perp}|^{\D-\hat{\D}}} C_{\hat{\cO}}(|x_{\perp}|^2 \partial^2_{\parallel}) \hat{\cO}(x^a),
\ee
with some differential operators $C_{\hat{\cO}}$. Here $\partial_\parallel$ is the derivative operator along the defect (in the $x^a$ directions).

A trivial example of a BOPE is obtained in the case in which
there is no actual defect. In this case, \eqref{qftbope} reduces essentially to a Taylor expansion.\footnote{With a caveat
that the operators $\hat{\cO}_n
= \partial^n_{x_{\perp}} \cO(x)|_{x_{\perp}=0}$ have to be diagonalized by adding terms of the form $c_i \partial_{x_{\perp}}^{n-2i}
(\partial_a \partial^a)^i \cO(x)|_{x_{\perp}=0}$. See \cite{Aharony:2003qf} for a detailed discussion.}

In the presence of a conformal defect, the one-point correlation function of a bulk operator and the two-point correlator of a bulk and a defect
operator are fixed by the unbroken defect conformal symmetry to be
\be
\< \cO(x)\> = \frac{a_{\cO}}{|x_{\perp}|^{\D}}, \qquad \< \cO(x) \hat{\cO}(0)\> =  \frac{b_{\cO \hat{\cO}} |x_{\perp}|^{\hat{\D} -\D}}{|x^{\m}|^{2\hat{\D}}},
\ee
with some constants $a_\cO$ and $b_{\cO \hat{\cO}}$, which in principle can be different on the two sides of the defect.

The two-point function of bulk operators is not fixed by symmetry alone and in general depends on one invariant crossratio $\xi$,
\be
\label{defxi}
\< \cO(x_1) {\cO}(x_2)\> = \frac{1}{|x_{1 \perp}|^{\D_1} |x_{2 \perp }|^{\D_2}} f(\xi), \qquad \xi = \frac{(x_1^a - x_2^a)^2 + (x_{1 \perp} - x_{2 \perp})^2}{4 x_{1 \perp} x_{2 \perp}}.
\ee
There are two interesting limits to consider. First, when $\xi \rightarrow 0$,
the two insertions approach each other but stay away from the defect. This limit is governed by the bulk OPE of the CFT without the defect.
We will focus on another limit, $\xi \rightarrow \infty$, i.e. when both insertions are close to the defect but remain apart from each other. This limit is governed by the BOPE:
\be
\label{bope}
\< \cO_1(x_1) {\cO}_2(x_2)\> \sim \sum_{\hat{\cO}} \frac{b_{\cO_1 \hat{\cO}} b_{\cO_2 \hat{\cO}}}{|x_{1 \perp}|^{\D_1 - \hat{\D}} |x_{2 \perp}|^{\D_2 - \hat{\D}}} C_{\hat{\cO}}(|x_{1 \perp}|^2 \partial^2_{1 \parallel}) C_{\hat{\cO}}(|x_{2 \perp}|^2 \partial^2_{2 \parallel}) \< \hat{\cO}(x_1^a) \hat{{\cO}}(x^a_2)\>.
\ee
Thus, we see that every $\hat{\cO}$ (primary or descendant) which appears in the BOPE of both $\cO_1$ and $\cO_2$ with the defect contributes to $f(\xi)$ as
\be
f(\xi) \sim \sum_{\hat{\cO}} b_{\cO_1 \hat{\cO}} b_{\cO_2 \hat{\cO}} \Big(\frac{4}{\xi}\Big)^{\hat{\D}}.
\ee
Therefore, from the $\xi \rightarrow \infty$ limit, we can read off the dimensions of operators $\hat{\cO}$ appearing in the BOPE of both $\cO_1$ and $\cO_2$. This provides very nontrivial information about the defect.

The contribution of descendants to \eqref{bope} can be resummed by solving the Casimir equation
(as for bulk conformal blocks). The result was obtained in \cite{McAvity:1995zd} (see also \cite{Liendo:2012hy, Billo:2016cpy}) and reads
\be
\label{boundaryblocks}
f(\hat{\D},\xi) = b^2_{\mathcal{O} \hat{\mathcal{O}}} \frac{1}{\xi^{\hat{\D}}} {}_2F_1
\Big(\hat{\D},\hat{\D}+1-d/2;2\hat{\D}+2-d;-\frac{1}{\xi}\Big)  
\ee
for identical scalar operators.
For later reference, we note that in two dimensions the hypergeometric function reduces to the Legendre function of the second kind \cite{bateman1953higher}
\be
{}_2F_1\Big(\hat{\D},\hat{\D}+1-d/2;2\hat{\D}+2-d;-\frac{1}{\xi}\Big) \, \stackrel{d=2}{=}\, \frac{2^{2 \hat{\D}}}{\sqrt{\pi}} \frac{\G[\hat{\D}+1/2]}{\G[\hat{\D}]} \xi^{\hat{\D}} Q_{\hat{\D}-1}(2\xi+1).
\ee

\section{Supergravity solution \label{solution}}

We now review the supersymmetric Janus solutions constructed
in \cite{Chiodaroli:2009yw}. The starting point is  type IIB supergravity compactified
on the four-torus $T^4$ or on the $K3$ manifold. In this paper, we will focus on the simpler case
in which the compactification manifold is a four torus.
We consider solutions with asymptotic geometries which are locally $AdS_3 \times S^3 \times T^4$.
Their holographic duals are marginal deformations of the two-dimensional $\mathcal{N}=(4,4)$ SCFT which arises as the IR fixed-point of the
worldvolume theory of the $D1/D5$ system \cite{Witten:1997yu,Vafa:1995bm,Dijkgraaf:1998gf}
(see ref. \cite{David:2002wn} for a review).
This CFT can be described more explicitly as the two-dimensional
sigma-model with the symmetric-orbifold target-space
$(T^4)^{Q_{D1} Q_{D5}}/S_{Q_{D1} Q_{D5}}$. The sigma-model is associated to the free Lagrangian
\begin{equation}
 S = {1 \over 2} \int \partial x^{i}_A \bar \partial x^{i}_A  -   \psi^i_A(z) \bar \partial \psi^i_A(z) -
\tilde \psi^i_A(\bar z)  \partial \tilde \psi^i_A (\bar z) \ ,
\end{equation}
where the index $i=1,\ldots,4$ runs over the coordinates of the four-torus and
the symmetric group acts by permuting the indices  $A$  of the $Q_{D1} Q_{D5}$ copies of the four-torus.

The particular Janus solutions under consideration are supported by the three- and five-form fluxes
and are dual to conformal interfaces obtained by marginally deforming the above CFT.
In the supergravity solutions, the six-dimensional dilaton and a linear combination of the axion
with the four-form potential interpolate smoothly between two different asymptotic values on opposite sides of the interface.
The six-dimensional dilaton is dual to the volume of the four-torus of the CFT, while the axion/four-form potential linear combination
corresponds to the source
for the orbifold $\mathbb{Z}_2$  twist operator.
In the CFT marginal deformation dual to the Janus solution, these operators have a step-function profile jumping across the interface.

The ten-dimensional metric for the supersymmetric Janus solution is written as
\bea
ds_{10}^{2} &=& f_{1,10}^{2 } ds^{2}_{AdS_{2}} + f_{2,10}^{2}ds^{2}_{S^{2}}  + \rho^{2}_{10}dw  d\bar w  + f_{3,10}^{2}ds^{2}_{T^{4}}  \ ,
\eea
where the metric factors $f_{1,10},f_{2,10},f_{3,10}$ and $\rho_{10}$ depend on the coordinates $w, \bar w$
of a two-dimensional Riemann surface $\Sigma$ with boundary.
The fields and metric factors have expressions in terms of two harmonic functions, $H$ and $K$, and two meromorphic functions, $A$ and $B$.
The harmonic functions $H, K, A+ \bar A$ and $B + \bar B$ all obey vanishing Dirichlet boundary conditions on the boundary $\partial \Sigma$.

Global regularity requires that the harmonic functions have isolated singular points on the boundary of $\Sigma$ and
imposes conditions on zeros and singularities of the relevant functions.
In particular, the singular points of $H$ correspond to asymptotic regions in which the geometry is locally
$AdS_3 \times S^3$. Aside from the Janus solutions that are the main focus of the present paper,
string-junction solutions with more than two asymptotic regions have been studied in refs. \cite{Chiodaroli:2009yw,Chiodaroli:2010mv}
and found to be dual to CFTs defined on star graphs.

The ten dimensional solutions solve the field equations
\begin{eqnarray}
&& \hskip -1.4cm \nabla^\mu P_\mu - 2i Q^\mu P_\mu + {1 \over 24} G_{\mu \nu \rho} G^{\mu \nu \rho} = 0 \ , \label{eomfirst}\\
&& \hskip -1.4cm \nabla^\rho G_{\mu \nu \rho} - i Q^\rho G_{\mu \nu \rho} - P^\rho \bar G_{\mu \nu \rho} + {2 \over 3} i F_{(5)\mu \nu \rho\sigma\lambda} G^{\rho \sigma \lambda} = 0\ , \\
&& \hskip -1.4cm R_\mu^{\ \nu} - P_\mu \bar P^\nu - \bar P_\mu P^\nu - {1 \over 6} (F^2_{(5)})^{\ \nu}_\mu
- {1 \over 8} (G_{\mu\rho \sigma} \bar G^{\nu \rho\sigma} + \bar G_{\mu\rho \sigma}  G^{\nu \rho\sigma}) + {1 \over 48} \delta_\mu^\nu G_{\rho\sigma\lambda}
\bar G^{\rho \sigma \lambda}  = 0  , \label{eomlast}
\end{eqnarray}
together with the Bianchi identities
\begin{eqnarray}
&&dP - 2 i Q \wedge P = 0 \ ,\\
&&dQ + i P \wedge \bar P = 0 \ ,\\
&&dG - i Q \wedge G + P \wedge \bar G = 0\ , \\
&&dF_{(5)} -  {i \over 8} G \wedge \bar G = 0 \ ,
\end{eqnarray}
and the self-duality condition for the five-form field strength $F_{(5)}$. In the above expressions, $P$ and $Q$ are a composite
one-form field strength and connection which can be expressed in terms of the axion $\chi$ and dilaton $\phi$ as
\begin{equation}
P = {1\over 2} \Big( d\phi + i e^\phi d\chi \Big) \ , \qquad Q = -{1\over 2} e^\phi d\chi  \ .
\end{equation}
Similarly, the complex three-form field strength $G$ is given in terms of the R-R and NS-NS field strengths $F_{(3)}$ and $H_{(3)}$ as
\begin{equation}
G = e^{- \phi/2} H_{(3)} + i e^{\phi/2} \big(F_{(3)} - \chi H_{(3)} \big).
\end{equation}

A key feature of the Janus solutions is  that the  three-form and five-form fluxes
are related to the volume forms of the unit $S^2$, $AdS_2$ and $T^4$ fibres $\hat \omega_{S^2},\hat \omega_{AdS_2}$ and $\hat \omega_{T^4}$ as
\bea F_{(3)} &=&dc^{(1)} \wedge \hat \omega_{AdS_2} + dc^{(2)} \wedge \hat \omega_{S^2} \ , \\
H_{(3)} &=&  db^{(1)} \wedge \hat \omega_{AdS_2} +  db^{(2)} \wedge \hat \omega_{S^2} \ , \\
F_{(5)} &=&  dC_K \wedge \hat \omega_{AdS_2} \wedge \hat \omega_{S^2} +   d\tilde C_K \wedge \hat \omega_{T^4} \ ,
\eea
where expressions for the scalar potentials $c^{(1,2)}, b^{(1,2)}, C_K$ and $\tilde C_K$ in terms of the relevant
harmonic functions  are given in ref. \cite{Chiodaroli:2009yw}.
Note that this structure is dictated by the $SO(2,1)\times SO(3)$ symmetry imposed in constructing the solutions.
For the reader's convenience, non-zero components of the various tensor fields are listed in table \ref{tab1}.
\begin{table}[t]
\begin{center}
\begin{tabular}{c||cc|cc|cc|cccc}
 & \multicolumn{2}{|c|}{$AdS_2$} & \multicolumn{2}{|c|}{$S^2$} & \multicolumn{2}{|c|}{$\Sigma$} & \multicolumn{4}{|c}{$T^4$} \\
 & $t$ \ & $z$ \ & $\theta$ \ & $\phi$ \ & $x$ \ & $y$ \ & $u_1$ \ & $u_2$ \ & $u_3$ \ & $u_4$ \ \\
 \hline
\multirow{2}{1cm}{$P, Q$} & & & & & X \ & & & & & \\
       & & & & &  & \ X \ & & & & \\
\hline
\multirow{4}{2cm}{$H_{(3)},F_{(3)}$} & \ X \ & \ X \ & & & \ X \ & & & & & \\
    & \ X \ & \ X \ & & &  & \ X \ & & & & \\
    &   &   & \ X \ & \ X \ & \ X \ & & & & & \\
    &   &   & \ X \ & \ X \ &  &  \ X \ & & & & \\
\hline
\multirow{4}{1cm}{$F_{(5)}$} & \ X \ & \ X \ & \ X \ & \ X \ & \ X \ & & & & & \\
    & \ X \ & \ X \ & \ X \ & \ X \ &  & \ X \ & & & & \\
    &   &   & & & \ X \ & & \ X \ & \ X \ & \ X \ & \ X \ \\
    &   &   & & &  & \ X \  & \ X \ & \ X \ & \ X \ & \ X \ \\
\end{tabular}\small
\caption{\small List of nonzero components of the five- and three-form fluxes and of the axion-dilaton composite field strength and connection
for general ten-dimensional Janus and
string-junction solutions.\label{tab1}}
\end{center}
\end{table}

It is convenient to reduce the solutions to six dimensions and write the metric in the Einstein frame
as
\bea
ds_6^{2} &=& f_{1}^{2 } ds^{2}_{AdS_{2}} + f^{2}_{2}ds^{2}_{S^{2}}  + \rho^{2 }dz  d\bar z \ ,
\eea
where $f^2_{1} = f^2_{1,10} f^2_{3,10} $, $f^2_{2} = f^2_{2,10} f^2_{3,10}$ and $\rho^2 = \rho^2_{10} f^2_{3,10}$.
Note that large classes of Janus and string-junction solutions have also been obtained directly
in six-dimensional supergravity in refs. \cite{Chiodaroli:2011nr,Chiodaroli:2011fn,Chiodaroli:2012vc}.
Here  we started with simple ten-dimensional
solutions and reduced them to six dimensions to better illustrate the top-down origin of our scalar deformations.
A useful identity involving the metric factors,
\be f^2_1 f^2_2 = H^2 \ , \ee
permits us to write the square root of the determinant of the metric as
\be \sqrt{-g}= \rho^2 H^2 \ . \ee

\begin{figure}[t]
\includegraphics[width=0.98\textwidth]{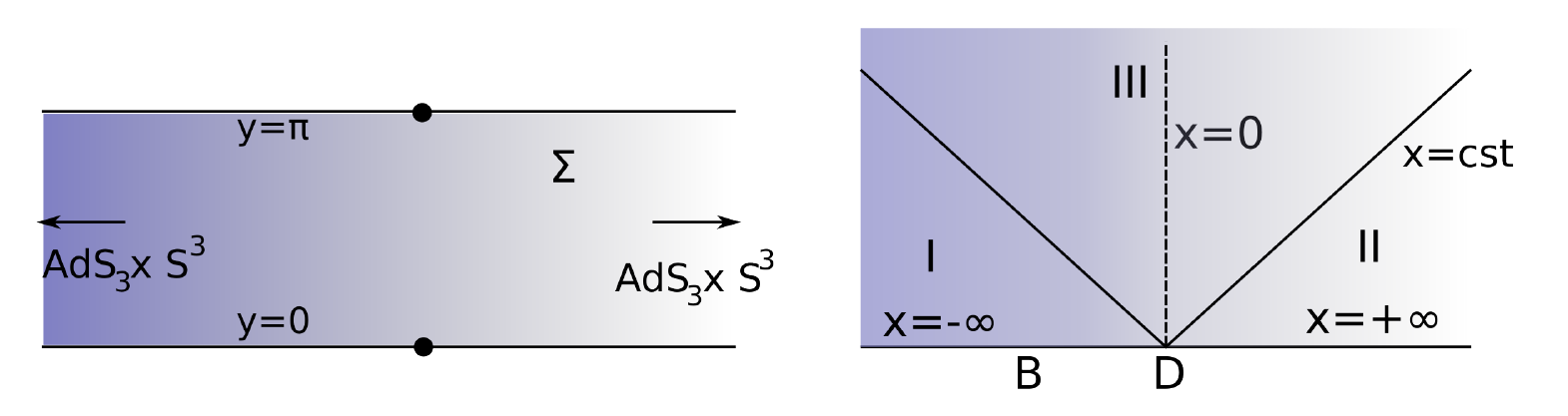}\small
\caption{\small Structure of the Janus solutions on the $x$-$y$ plane and $x$-$z$ plane.
The Riemann surface $\Sigma$ is the infinite strip. For a solution supported only by RR fluxes, the functions $A,B$ and $K$ have singularities on
both boundaries  at $x=0$. The spacetime boundary B is reached for $x \rightarrow \pm \infty$ and
for $z \rightarrow 0$ on the $AdS$ slice. The latter limit corresponds to the location of the defect D.
\label{fig1}}
\end{figure}

\begin{figure}
\includegraphics[width=0.48\textwidth]{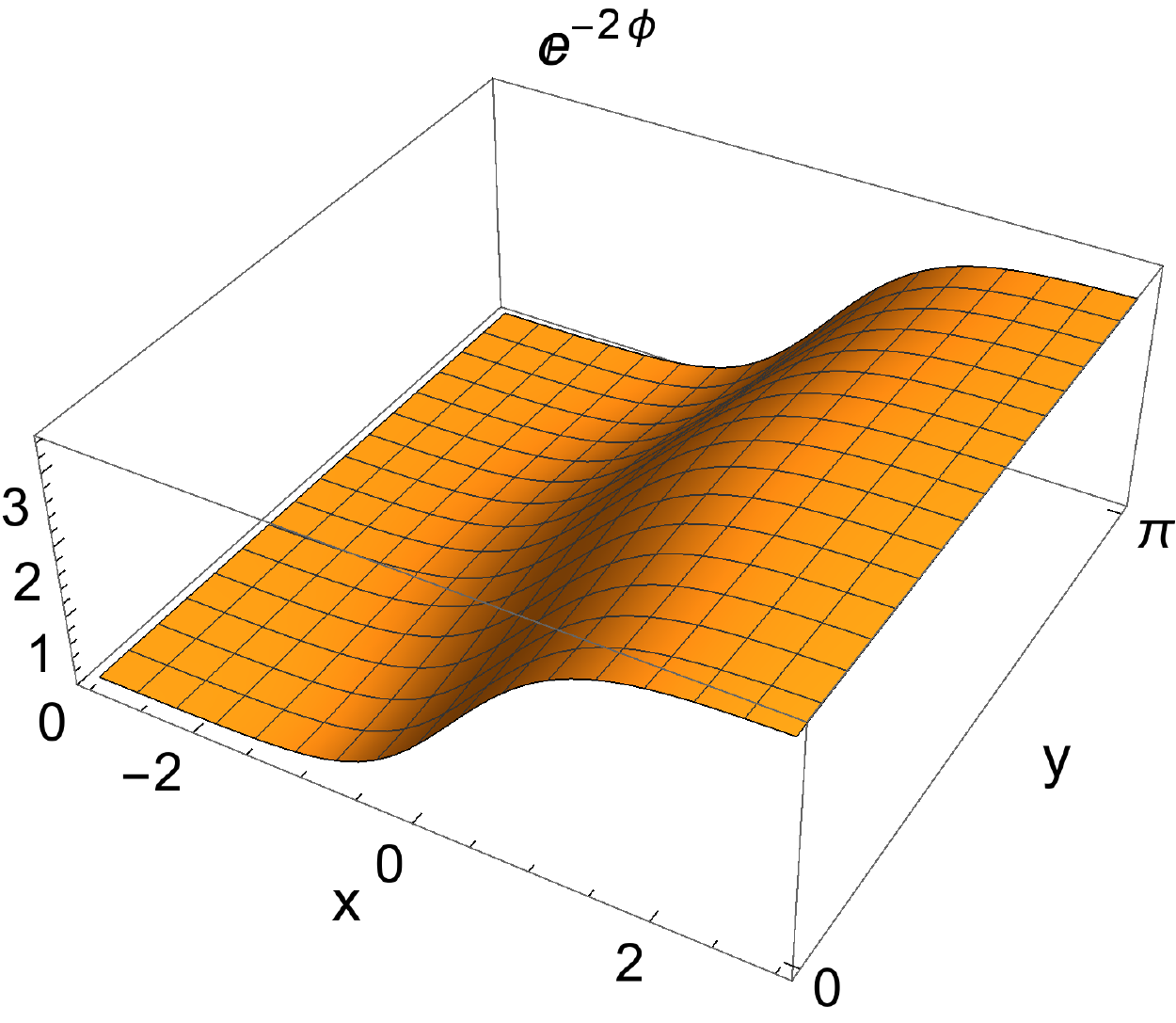} \ \
\includegraphics[width=0.48\textwidth]{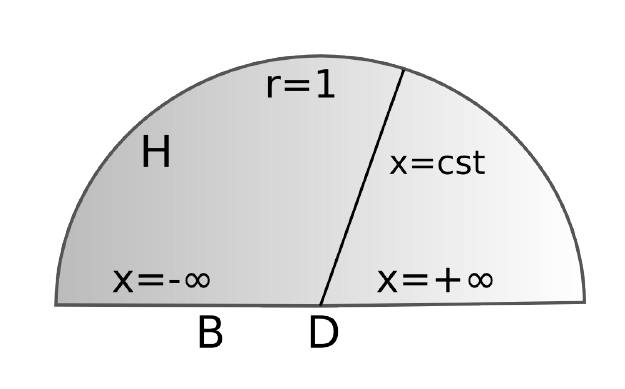} \small
 \caption{
 \small Dilaton profile for the supersymmetric Janus solution with $\psi=1,\theta=0$ in the $x-y$ plane (left)
 and sketch of the finite-temperature solution (right). The black-hole horizon H is located at $r=1$, the defect D is located at $r=0$.  \label{fig2}}
\end{figure}

We shall now focus on the simplest solutions  which have two asymptotic $AdS_3 \times S^3$
regions, a simply-connected Riemann surface, and are supported only by R-R charges (see ref. \cite{Chiodaroli:2010mv} for a discussion of the Page charges associated to our solutions).
We  take the infinite strip with coordinates $x,y$ as the Riemann surface $\Sigma$.
Figure \ref{fig1} displays the pole configuration for the solutions of interest in the $x$-$y$ plane,
as well as a sketch of the $x$-$z$ plane which illustrates the position of the defect in our coordinates.
The two asymptotically $AdS_3$ regions are approached as $x \rightarrow \pm \infty$.
Nontrivial Janus solutions are parameterized by two parameters, $\psi$ and $\theta$, associated to the jump in the dilaton and axion respectively.
The undeformed case is obtained by $\theta=0=\psi$. A plot of the ten-dimensional dilaton for the solution with nonzero $\psi$ can be found in figure \ref{fig2}.

The following simple expressions for the harmonic function
$H$ and the ratios of metric factors \cite{Chiodaroli:2009yw,Chiodaroli:2010ur},
\begin{align}
&H = 2 \hat L \cosh x \sin y \ ,&
&{\rho^2 \over f^2_1 } = {\kappa^2 \over  \cosh^2 x }\ ,&
&{\rho^2\over f_2^2} = {1\over \sin^2 y} + {\kappa^2- 1 \over  \cosh^2 x  }\ ,& \label{relmetr3}
\end{align}
will be useful in the remainder of the paper.
We have introduced the interface parameter $\kappa = \cosh\psi \cosh\theta$ and denoted with $x,y$
real and imaginary part of the complex coordinate of $\Sigma$.

So far, we have discussed zero-temperature solutions. However, as explained in ref. \cite{Estes:2015jha}, a black-hole
solution can be obtained by replacing the
$AdS_2$ fibre with the ``black -hole slice" metric
\be ds^2_{AdS_2} \quad \mapsto \quad ds^2_{BH} = -  {1-r^2 \over r^2} (2 \pi T)^2 dt^2 + {dr^2 \over r^2 ( 1 - r^2) } \ , \qquad 0 \leq r \leq 1 \ ,  \ee
where we have introduced the explicit temperature dependence as $T$ and we
have employed the coordinate transformation $\rho = 1/r$ as compared to ref. \cite{Estes:2015jha}.
For these solutions, the interface is located at $r=0$ and the horizon at $r=1$. In the absence of an interface,
the Fefferman-Graham coordinate expansion close to the boundary
reproduces the one of the BTZ black hole order by order in the radial parameter.
Black-hole solutions obtained with this technique are dual to finite-temperature CFTs defined on the real line times the thermal circle.\footnote{
We would like to emphasize that additional global
identifications, analogous to the ones arising for the BTZ black-hole solution in Poincar\'{e} coordinates,
are required to reproduce the global structure of a solution dual  to  a finite-temperature CFT
defined on a two-torus.} The structure of the solution on the $r-x$ plane can be found in figure \ref{fig2}.

\bigskip

We are now ready to discuss
the ten-dimensional origin of the probe scalars obeying the massless Klein-Gordon equation which we will solve in the next section.
Given the general structure of our interface solutions, we consider simple deformations of the supersymmetric Janus geometry
which leave invariant the volume forms of the fibres.

The simplest of such deformations affects only the components of the metric along the $T^4$ fibre and has the form
\be g_{ij} = f^2_{3,10} \text{exp} (  t^a_{ij} \Phi^a ) = f^2_{3,10} \big( \delta_{ij} +
\Phi^a t^a_{ij} + \ldots \big) , \qquad  (t^a)_{i i} = 0 \ , \qquad i,j = 6,7,8,9 \ , \label{deftorus}  \ee
where $t^a_{ij}$ ($a=1,\ldots,9$) are constant symmetric-traceless tensors along the compact $T^4$ directions.

A deformation of this form leaves the three-form and five-form fields invariant, as well as the axion and dilaton. Inspecting the
equations of motion (\ref{eomfirst})-(\ref{eomlast}), it is immediate to verify that the  only non-trivial equation  is (\ref{eomlast}),
which reduces to the equation for a free scalar,\footnote{To obtain the equation for the free scalar (\ref{freescalar}),
it is simpler to consider the field equations
with mixed indices.}
\be  R_\mu^{\ \nu} = 0 \quad  \Rightarrow \quad  \nabla^2 \Phi^a = 0 \ . \label{freescalar} \ee
Alternatively, one can consider the $\Phi^a$ dependence of the action,
\be  S[\Phi] = \int d^4 x \sqrt{g} \Big(  - \partial_\mu \Phi^a \partial^\mu \Phi^a \Big)  \ . \ee
In the remainder of the paper, we will study the equation of motion for these free scalars in the Janus geometry and, in particular, find that
it is separable, allowing for an explicit analytic solution. As mentioned before,
one of the advantages of the top-down approach is that the dual
$\cN=(4,4)$ SCFT is known explicitly.
In particular, its operators are
classified using the $\cN=(4,4)$ superconformal algebra and are labelled by their representation of
the $SO(4)_I$ symmetry acting on the coordinates of the four-torus. Hence, the operator dual to our scalar
deformation is identified as
\begin{displaymath}
 {\cal O}^{\{ij\} } = \partial x^{\{ i}_A \bar \partial x^{j\}}_A = \partial x^{ i}_A \bar \partial x^{j}_A +
 \partial x^{j}_A \bar \partial x^{i}_A - {1 \over 4} \delta^{ij} \partial x^{k}_A \bar \partial x^{k}_A \ .
\end{displaymath}

The reader might wonder if considering deformations leaving the fibres' volume forms invariant
might lead to other simple perturbations.  In particular,
a potentially interesting deformation involves the metric on the $AdS_2$ slice,
\be g_{\mu \nu} = \bar g_{\mu \nu} + h_{\mu \nu}  , \qquad  h_{\mu \nu} \bar g^{\mu \nu}  = 0 \ , \qquad \mu ,\nu = 0,1 \ ,  \ee
and leaves all other fields invariant.
Repeating the argument above, we obtain
\be  R_\mu^{\ \nu} \big|_{{\cal O} (h)} = 0 \quad \Rightarrow \quad
\nabla^2 h_{\mu \nu} - \nabla_\mu \nabla^\rho h_{\nu \rho} - \nabla^\nu \nabla^\rho h_{\mu \rho} + 2 \bar R^{\rho \ \sigma}_{\ \mu \ \nu} h_{\rho \sigma} = 0 \ , \ee
where $\bar R^{\rho \ \sigma}_{\ \mu \ \nu}$ is the Riemann tensor of the background geometry. These are three differential equations in two functions $h_{tt}$ and $h_{tz}$
and  force a trivial dependence on the $t,z$ coordinates for $h_{tt}$ and $h_{tz}$.
Hence,
to have non-trivial solutions, one needs to turn on the trace part of the $AdS_2$ metric perturbation.
While deformations of this sort are very interesting, their treatment is considerably more involved and will be deferred to future work.

\section{Solving the massless Klein-Gordon equation} \label{linearmodes}

From now on, we drop the index parametrizing different deformations on the torus and work with the single scalar $\Phi$.
The massless Klein-Gordon equation \eqref{freescalar} for the Janus geometry with two $AdS_3$ asymptotic regions can then be written as
\be {1 \over \sqrt{-g}} \partial_\mu (\sqrt{-g} g^{\mu \nu} \partial_\nu \Phi) =
{1 \over \rho^2} \left\{ { \partial_a (H^2 \partial_a ) \over H^2} + {\rho^2 \over f^2_1} \nabla^2_{AdS_2}
+{\rho^2 \over f^2_2} \nabla^2_{S^2}  \right\} \Phi = 0 , \qquad a=x,y  \ee
where $\nabla^2_{S^2}$ is the Laplacian on the unit two-sphere and $\nabla^2_{AdS_2}$ is the Laplacian on the unit $AdS_2$ slice,
\be \nabla^2_{AdS_2} = z^{2} \big( \partial^2_z -\partial_t^2 \big) \ . \ee
In case of the black-hole solution, we replace
\be \nabla^2_{AdS_2} \quad \mapsto  \quad \nabla^2_{BH} = r^2 \Big(  \partial_r (1 -r^2) \partial_r - \frac{1}{(2 \pi T)^2} {\partial_t^2 \over 1 - r^2} \Big) \ .    \ee
For the massless scalar field, the Klein-Gordon equation is separable. In this section we will present the solution.

First, we note that the differential equations on the two-dimensional slices,
\bea \label{AdSLaplace} \nabla^2_{AdS_2} f = \nu(\nu+1) f \ , \qquad \nabla^2_{BH} f = \nu(\nu+1)  f \ , \eea
admit solutions in terms of Bessel and associated Legendre functions,
\bea {\rm AdS}: && f(z,t) = C_1 \sqrt{z} J_{\nu + \frac{1}{2}} (\omega z)e^{\pm i \omega t} + C_2 \sqrt{z} Y_{\n + \frac{1}{2}} (\omega z)e^{\pm i \omega t} \ , \\
{\rm BH}: && f(r,t) = C_1 P^{\frac{i \omega}{2\pi T}}_\nu \Big({1 \over r} \Big)e^{\pm i \omega t} + C_2 Q^{\frac{i \omega}{2 \pi T}}_\nu \Big({1 \over r} \Big) \ . \eea
In the Euclidean signature, the Bessel functions $J_{\n+1/2}$ and $Y_{\n+1/2}$ should be replaced by the modified Bessel functions $K_{\n+1/2}$ and $I_{\n+1/2}$.

The effective mass on the $AdS_2$ slice is related to the dimension of the defect operator simply by $\n = \hat{\D}-1$. We will later
fix the possible values of $\n$ by imposing the absence of sources on the boundary. This will result in a discrete tower of possible masses on the $AdS_2$ slice and
determine the dimensions of the operators appearing in the BOPE of the dual operator with the defect. To extract the BOPE coefficients themselves, we need to compute the two-point correlation function of bulk operators.

The Laplace operator on the two-sphere has the usual spherical harmonics as its eigenfunctions. Thus the probe scalar can be expanded as
\bea \hskip -0.5cm {\rm AdS}: && \hskip -0.5cm
\Phi = \sum \Big( \Phi^{(1)}_{\nu \omega  l m}(x,y) \sqrt{z} J_{\nu + \frac{1}{2}} (\omega z) + \Phi^{(2)}_{\nu \omega  l m}(x,y) \sqrt{z} Y_{\nu + \frac{1}{2}} (\omega z)  \Big) e^{\pm i \omega t}
Y_{lm}(\theta, \phi) , \\
\hskip -0.5cm {\rm BH}: && \hskip -0.5cm
\Phi = \sum \Big( \Phi^{(1)}_{\nu \omega  l m}(x,y) P^{\frac{i \omega}{2\pi T}}_{\nu} \Big({1 \over r} \Big)  + \Phi^{(2)}_{\nu \omega  l m}(x,y) Q^{\frac{i \omega}{2\pi T}}_{\nu} \Big({1 \over r} \Big)  \Big) e^{\pm i \omega t}
Y_{lm}(\theta, \phi)  , \quad  \eea
where the spherical harmonics satisfy $\nabla^2_{S^2} Y_{lm}(\theta, \phi)= - l (l+1)  Y_{lm}(\theta, \phi)$.
In both cases, the Klein-Gordon equation becomes
\bea &&  \left\{ { \partial_a (H^2 \partial_a)  \over H^2} + \nu(\nu+1) {\rho^2 \over f^2_1} - l (l+1) {\rho^2 \over f^2_2}
\right\} \Phi^{(i)}_{\nu \omega l m } = 0 \ ,
\eea
or using \eqref{relmetr3}
\be \left\{ { \partial_x \cosh^2 x \ \partial_x + \alpha(\alpha+1) \over \cosh^2 x } + { \partial_y \sin^2 (y) \ \partial_y - l (l+1)  \over \sin^2 (y)}
\right\} \Phi^{(i)}_{\nu \omega l m } = 0 ,  \qquad i=1,2 \ , \label{eqsep} \ee
where the constant $\alpha$ is determined in terms of  $l$ and $\nu$ by
\bea
\label{defalpha}
\alpha (\alpha+1) &=& l(l+1) + \kappa^2 \Big( \nu(\nu+1) - l(l+1) \Big) \ . \eea
Equation (\ref{eqsep}) is separable. Moreover, the $y$-dependent part combines with the Laplacian on the $S^2$
slice to give the Laplacian on the three-sphere, with eigenvalues $-k(k+2)$.\footnote{We introduce ${\cal Y}_{klm}(y,\theta,\phi)$, which satisfies $\nabla^2_{S^2} {\cal Y}_{klm}= - l (l+1)  {\cal Y}_{klm}$ and $[\partial_y (\sin^2 (y) \ \partial_y) - l (l+1)] {\cal Y}_{klm} = - k(k+2) \sin^2 (y) {\cal Y}_{klm}$.}

To get a better intuition, it is useful to represent the field equation in the form
\be
\label{3deq}
\Box_{3d}\Phi_{l,k} = \Big(\frac{l(l+1)}{\cosh^2 x}(\kappa^2-1)+ k(k+2)\Big) \Phi_{l,k} = M^2_{3d}(x) \Phi_{l,k},
\ee
where $\Box_{3d}$ is the Laplacian on the three-dimensional asymptotically locally-$AdS_3$ space with the metric
\be
\label{3dmetric}
ds_{3d}^2 =dx^2 +  \frac{\cosh^2 x}{\kappa^2} ds^2_{AdS_2}= dx^2 +  \frac{\cosh^2 x}{\kappa^2} \frac{dz^2 - dt^2}{z^2}.
\ee
Thus the presence of the defect does not just modify the $AdS_3$ metric,
but also introduces a position-dependent mass, $M^2_{3d}(x)$.
Without the defect, the dimension of the dual operator is simply $\D=k+2$. In the presence of the defect,
this operator gets decomposed into a tower of operators with dimensions $\hat{\D}_n = \n(n)+1$,
the precise form of which we will deduce later.
Importantly, the dimensions of the members of this tower are not related in a simple way to the dimension of the original operator,
in contrast to the case in which the defect is absent. This is a characteristic feature of the BOPE.
In the process of bringing an operator close to the defect, divergences can appear and regularization is required as in the case of composite operators.

After solving the $AdS_2$ and $S^3$ parts of the problem, we are left with a single ordinary differential equation,
\be  { \partial_x \cosh^2 x \ \partial_x + \alpha(\alpha+1) \over \cosh^2 x }\chi^{(i)}_{ \omega \nu k l m }(x) = k(k+2) \chi^{(i)}_{ \omega \nu k l m }(x),
\qquad i=1,2 \ ,
\ee
which is solved by
\bea \chi^{(i)}_{ \omega \nu k l m }(x) &=& {  c_3 P^{k+1}_{\alpha}(\tanh x ) +
c_4  Q^{k+1}_{\alpha}(\tanh x) \over \cosh x} \label{sol-x} \ . \eea
To summarize, there are eight classes of solutions for the probe scalar in the Janus background. For the AdS slicing we have:
\bea {\rm AdS}:
&& \Phi^{(1)}_{\omega \nu k l m} = {\sqrt{z} \, \omega^{-(\nu + \frac{1}{2})} \over \cosh x} P^{k+1}_{\alpha} \big( \tanh x \big) J_{\nu + \frac{1}{2}} \big( \omega z \big) e^{\pm i \omega t} {\cal Y}_{klm}
\big( y, \theta, \phi \big)  \ , \no \\
&& \Phi^{(2)}_{\omega \nu k l m} = {\sqrt{z} \, \omega^{(\nu + \frac{3}{2})} \over \cosh x} P^{k+1}_{\alpha} \big( \tanh x \big) Y_{\nu + \frac{1}{2}} \big( \omega z \big) e^{\pm i \omega t} {\cal Y}_{klm}
\big( y, \theta, \phi \big)\no\ , \\
&& \Phi^{(3)}_{\omega \nu k l m} = {\sqrt{z} \, \omega^{-(\nu + \frac{1}{2})} \over \cosh x} Q^{k+1}_{\alpha} \big( \tanh x \big) J_{\nu + \frac{1}{2}} \big( \omega z \big) e^{\pm i \omega t} {\cal Y}_{klm}
\big( y, \theta, \phi \big) \no \ , \\
&& \Phi^{(4)}_{\omega \nu k l m} = {\sqrt{z} \, \omega^{(\nu + \frac{3}{2})} \over \cosh x} Q^{k+1}_{\alpha} \big( \tanh x \big) Y_{\nu + \frac{1}{2}} \big( \omega z \big) e^{\pm i \omega t} {\cal Y}_{klm}
\big( y, \theta, \phi \big)\ , \label{modes} \eea
where ${\cal Y}_{klm}$ are the three-sphere spherical harmonics and $\omega \geq 0$. The factors of $\omega$ have been included so
that the limit $\omega \rightarrow 0$
can be taken smoothly.  For the black-hole solutions we have
\bea {\rm BH}: && \Phi^{(1)}_{\omega \nu k l m} = {1 \over \cosh x} P^{k+1}_{\alpha} \big( \tanh x \big) P^{\frac{i \omega}{2\pi T}}_{\nu} \Big({1 \over r}  \Big) e^{\pm i \omega t}
{\cal Y}_{klm}
\big( y, \theta, \phi \big)  \ , \no \\
&& \Phi^{(2)}_{\omega \nu k l m} = {1 \over \cosh x} P^{k+1}_{\alpha} \big( \tanh x \big) Q^{\frac{i \omega}{2\pi T}}_{\nu} \Big({1 \over r}  \Big) e^{\pm i \omega t} {\cal Y}_{klm}
\big( y, \theta, \phi \big)\no\ , \no \\
&& \Phi^{(3)}_{\omega \nu k l m} = {1 \over \cosh x} Q^{k+1}_{\alpha} \big( \tanh x \big) P^{\frac{i \omega}{2\pi T}}_{\nu} \Big({1 \over r}  \Big) e^{\pm i \omega t} {\cal Y}_{klm}
\big( y, \theta, \phi \big) \no \ , \no \\
&& \Phi^{(4)}_{\omega \nu k l m} = {1 \over \cosh x} Q^{k+1}_{\alpha} \big( \tanh x \big) Q^{\frac{i \omega}{2\pi T}}_{\nu} \Big({1 \over r}  \Big) e^{\pm i \omega t} {\cal Y}_{klm}
\big( y, \theta, \phi \big)\ . \label{modesBH} \eea
Note that at this stage the parameters $k,l,m$ are quantized. The general solution is obtained by superimposing these modes.

The defect modes correspond to those solutions which vanish at the boundary away from the defect, i.e. when $x \rightarrow \pm \infty$.
Using simple properties of Legendre functions,
we find that this requirement is satisfied only if $\a = k + 1 + n$ for a non-negative integer $n$. Note that \eqref{defalpha} tells us what discrete
values the $\n$ parameter can take,
\begin{equation}
 \nu(n) = \Big\{ {1 \over 4} + {(k+n+1)(k+n+2)\over \kappa^2} +  l(l+1) \Big( 1 - {1 \over \kappa^2} \Big) \Big\}^{1\over 2} -{1 \over 2} \ ,
\end{equation}
where we have taken the positive solution to (\ref{defalpha}).

From \eqref{AdSLaplace} we know that the effective mass on the $AdS_2$ slice is $m^2 = \nu(\nu+1) = \hat{\Delta}_n (\hat{\Delta}_n-1)$.
Thus, for the defect modes, we obtain the conformal dimension of the corresponding fields:
\be
m_{l,n}^2 = \hat{\Delta}_n (\hat{\Delta}_n-1) = l(l+1)+ \frac{1}{\k^2} \Big[(k+n+2)(k+n+1) - l(l+1) \Big],
\ee
where $n = 0,1,2,3,...$ is an integer and we have set $\alpha = k+1+n$.  Note that the massless $AdS_2$ scalar cannot appear in the spectrum.

As a simple check, consider the pure $AdS_3$ scenario, i.e. assume that the dilaton and axion do not jump across the defect ($\k=1$).
Then for $k=0$, we have
\begin{align}
&m_{n}^2 = (n+2)(n+1),&
&\Rightarrow&
&\hat{\Delta}_n = n + 2&
\end{align}
The higher values of
$n$ then give the dimension of the higher-order operators appearing in the expansion.\footnote{Consider the expansion ${\cal O}(x_\perp) =  \sum_{n=1}^\infty {\cal O}^{(n)} x_\perp^{n-1}$.
The dimension of ${\cal O}^{(0)}$ would be the same dimension as ${\cal O}$, which is two for a marginal operator.
The dimension of the higher order terms is then $[{\cal O}^{(n)}] = \Delta_n = n+1$. }
For higher values of $k$, we have $\Delta_n = k+2+n$,  in agreement with the fact that the $AdS_3$ mass is given by $m^2_{AdS_3} = k(k+2)$.

\section{Bulk-to-bulk propagator and  two-point correlation function} \label{propcorr}

In this section, we use the linearized modes to construct the bulk-to-bulk propagator satisfying
vanishing boundary conditions at the boundary of the spacetime and infalling boundary conditions at the Poincar\'{e}
horizon. The latter choice is appropriate for the retarded holographic propagator. In Euclidean signature, it is sufficient
to impose regularity in the spacetime's interior.
The propagator can be constructed as a summation over modes which satisfy
prescribed boundary conditions. The modes which vanish at the boundary (away from the defect) must necessarily have $\alpha \in \mathbb{N}$. This gives the quantization condition for the masses on the $AdS_2$ slice and hence the dimensions of operators appearing in the BOPE of the dual operator with the defect.

In what follows, we will effectively ignore the $S^3$ part of the geometry and construct propagators directly for the equation \eqref{3deq}.
\subsection{Bulk-to-bulk propagator}
As we will show shortly, the bulk-to-bulk propagator is given explicitly by
\begin{align}
\label{bulktobulkprop}
G_{\D} =& \frac{1}{2 \pi} \int\displaylimits_{0}^{\infty}d \o e^{- i \o(t-t')} \sum_{n=0}^{\infty}\frac{(n+\D-1/2)n!}{(n+2\D-2)!} \tilde{G}_{\n+1/2}[z,z']
\no \\ & \qquad \qquad \qquad \qquad \qquad \times
\frac{P^{\D-1}_{n+\D-1}(\tanh x)}{\cosh x} \frac{P^{\D-1}_{n+\D-1}(\tanh x')}{\cosh x'}
+ \text{c.c.},
\end{align}
where
\begin{align}
\tilde{G}_{\n}[z,z'] =
\frac{1}{W[\sqrt{z} H_{\n}^{(1)}(|\o| z),\sqrt{z} J_{\n}(|\o| z)]\Big|_{z=z'}} \Big[&\theta(z-z') \sqrt{z} H^{(1)}_{\n}(|\o| z) \sqrt{z'} J_{\n}(|\o| z') +\no \\
&\theta(z'-z) \sqrt{z} J_{\n}(|\o| z) \sqrt{z'} H^{(1)}_{\n}(|\o| z')  \Big]
\end{align}
is the propagator in the $z$ direction which satisfies the equation
\begin{align}
\label{zproblem}
\Big(z^2 \partial_z^2 +\o^2 z^2 -\n (\n+1) \Big) \tilde{G}_{\n+1/2}[z,z'] = z^2 \delta(z-z'),
\end{align}
vanishes as $z$ approaches zero and satisfies infalling boundary conditions at the horizon of the $AdS_2$ slice, i.e. as $z$
goes to infinity. $W[\sqrt{z} H_{\n}^{(1)}(|\o| z),\sqrt{z} J_{\n}(|\o| z)]$ denotes the Wronskian of the two functions, which,
for the case at hand,  is equal to $-2 i/\pi$. The Green's function is a continuous function with discontinuous
first derivative at $z=z'$.\footnote{To obtain the result in Euclidean signature we continue $\o^2 \rightarrow - \o^2$.
Hankel function $H_{\n}$ gets replaced by $K_{\n}$ (up to a simple prefactor)
which is regular as $z\rightarrow \infty$ and $J_{\n}$ gets replaced by $I_{\n}$.
The Wronskian is then $W[\sqrt{z} K_{\n}(|\o| z),\sqrt{z} I_{\n}(|\o| z)] =1$.}  Note that the choice of the Hankel function of the first
kind is determined by requiring the Green's function to satisfy ingoing boundary conditions at the horizon,
while the choice of the Bessel function of the first kind $J_{\n}(|\o| z)]$
is determined by requiring the bulk-to-bulk propagator to vanish at the boundary of the $AdS_2$ slice.
The quantization of $\alpha = n+\D-1 = n + k + 1$ is determined by requiring the bulk-to-bulk propagator to vanish as $x \rightarrow \pm \infty$.
This means that the bulk-to-bulk propagator is decomposable in terms of the defect modes.

Let us show by explicit computation that \eqref{bulktobulkprop} indeed gives the propagator.
Acting with the three-dimensional Laplacian on $G_{\D}$, we note that the non-vanishing contribution comes from the step-function
in the $z$ direction. Explicitly,
\begin{align}
\left( \Box_{3d} - M_{3d}^2 \right) G_{\D} =& \frac{1}{2 \pi} \int\displaylimits_{-\infty}^{\infty}d \o e^{- i \o(t-t')}
\bigg[ \frac{\kappa^2}{\cosh^2 x} z^2\delta(z-z')  \bigg]
\no \\& \qquad
\sum_{n=0}^{\infty}\frac{(n+\D-1/2)n!}{(n+2\D-2)!} \frac{P^{\D-1}_{n+\D-1}(\tanh x)}{\cosh x} \frac{P^{\D-1}_{n+\D-1}(\tanh x')}{\cosh x'}.
\end{align}
Next, we use the completeness relation for the associated Legendre polynomials
\be
\cosh x \cosh x' \delta(x-x') = \sum_{n=0}^{\infty}\frac{(n+\D-1/2)n!}{(n+2\D-2)!} P^{\D-1}_{n+\D-1}(\tanh x) P^{\D-1}_{n+\D-1}(\tanh x')
\ee
and obtain
\be
\left( \Box_{3d} - M_{3d}^2 \right) G_{\D} =\frac{1}{\sqrt{|g_{3d}}|} \delta(t-t') \delta(x-x')\delta(z-z')\,,
\ee
with $\sqrt{|g_{3d}}| = \cosh^2 x/(\k^2 z^2)$.
Thus, we see that the bulk-to-bulk propagator can be indeed represented as a sum of the modes derived before.

The Euclidean version is
\begin{align}
\label{Ebulktobulkprop}
G_{E,\D} = \frac{1}{2 \pi} \int\displaylimits_{-\infty}^{\infty}d \o & e^{- i \o(t-t')} \sum_{n=0}^{\infty}\frac{(n+\D-1/2)n!}{(n+2\D-2)!}
\nonumber \\& \qquad
\tilde{G}_{E,\n+1/2}[z,z']
\frac{P^{\D-1}_{n+\D-1}(\tanh x)}{\cosh x} \frac{P^{\D-1}_{n+\D-1}(\tanh x')}{\cosh x'} \,,
\end{align}
with
\begin{align}
\tilde{G}_{E,\n}[z,z'] = \frac{1}{W[\sqrt{z} K_\nu(|\o| z),\sqrt{z} I_\nu(|\o| z)]\Big|_{z=z'}} \no
\Big[&\theta(z-z') \sqrt{z} K_{\n}(|\o| z) \sqrt{z'} I_{\n}(|\o| z') +\\ &\theta(z'-z) \sqrt{z} I_\nu(|\o| z) \sqrt{z'} K_{\n}(|\o| z')  \Big]   \,.
\end{align}
In this case, we have $W[\sqrt{z} K_\nu(|\o| z),\sqrt{z} I_\nu(|\o| z)] = 1$.
The continuation back to Minkowski signature is achieved by replacing $J_\nu \rightarrow I_\nu$ and $K_\nu \rightarrow H^{(2)}_\nu$ for
$\omega < 0$ and $K_\nu \rightarrow H^{(1)}_\nu$ for $\omega > 0$.

Finally, for the finite-temperature case, we can proceed along the same lines and obtain the expression
\begin{align}
\tilde{G}_{E,T,\n}[z,z'] = \frac{1}{W[Q^{\frac{\omega}{2\pi T}}_{\nu} \Big({1 \over r}\Big),P^{\frac{\omega}{2\pi T}}_{\nu} \Big({1 \over r}  \Big)]\Big|_{r=r'}}
\Big[&\theta(r-r') Q^{\frac{\omega}{2\pi T}}_{\nu} \Big({1 \over r}\Big) P^{\frac{\omega}{2\pi T}}_{\nu} \Big({1 \over r'}  \Big) \no \\ & + \theta(r'-r) P^{\frac{\omega}{2\pi T}}_{\nu} \Big({1 \over r}  \Big) Q^{\frac{\omega}{2\pi T}}_{\nu} \Big({1 \over r'}\Big) \Big] \ .
\end{align}
This propagator is regular at the black-hole horizon at $r=1$
and is appropriate for the holographic dual of a CFT defined on the real line times the thermal circle.
Note that the boundary conditions need to be modified if we introduce additional global identifications in the black-hole geometry.

\subsection{Bulk-to-boundary propagator}
The bulk-to-boundary propagator $K_{\D}(x_0, \vec{x}, \vec{x}')$ can be obtained by taking the limit \cite{Giddings:1999qu, Papadimitriou:2004rz}
\begin{align}
K(z,t,x;t',x_\perp') = \lim_{u' \rightarrow 0} \frac{(2 \Delta - 2)}{(u')^\Delta} G_\Delta(z,t,x;u',t',x_\perp'),
\end{align}
where the $u'$ coordinate denotes the radial coordinate in the Poincar\'e patch of $AdS_3$. For the $AdS$-slicing coordinates we are employing, the  Fefferman-Graham
expansion needs to be computed separately on both sides of the defect (regions I and II in figure \ref{fig1}).
The leading-order terms are:
\begin{eqnarray}
\text{region I}:&& \qquad  \quad  x' \sim - \ln ( 2 u'/\kappa | x_\perp'|) \ , \qquad   z' \sim - x_\perp' \ , \\
\text{region II}: && \qquad \quad x' \sim + \ln ( 2 u'/\kappa |x_\perp'|)\ , \qquad  z' \sim + x_\perp'\ .
                      \end{eqnarray}
To compute the limit, we use that \cite{bateman1953higher}
\be
P_{n+\D-1}^{\D-1}(\tanh x') \sim  \frac{(-1)^{\D-1} 2^{-\frac{\D-1}{2}} \G(n+ 2\D-1) (1-\tanh x')^{\frac{\D-1}{2}}}{(\D-1)! \G(n+1)}, \quad \text{as} \quad x' \rightarrow \infty.
\ee
The other side of the defect is approached as $x' \rightarrow -\infty$, this would introduce a $(-1)^n$ prefactor.
For now we postpone this issue until later. A short computation gives
\begin{align}
G_{\D} \sim &\frac{1}{(\D-2)!}\Big(-\frac{1}{2} \Big)^{\D-1} \frac{1}{2(\D-1)\pi} \int\displaylimits_{-\infty}^{\infty}d \o e^{- i \o(t-t')} \times \nonumber \\ &\sum_{n=0}^{\infty} (n+\D-1/2)\tilde{G}_{\n+1/2}[z,z'] \frac{1}{(\cosh x')^{\D}} \frac{P^{\D-1}_{n+\D-1}(\tanh x)}{\cosh x} \quad \text{as} \quad x' \rightarrow \infty,
\end{align}
so that the bulk-to-boundary propagator can be read off:
\begin{align}
\label{bulkboundprop}
K_{\D}(z,t,x;t',x_{\perp}') = &\frac{1}{(\D-2)!}\Big(-\frac{1}{2} \Big)^{\D-1} \frac{1}{\pi} \frac{1}{\kappa^\Delta} \int\displaylimits_{-\infty}^{\infty}d \o e^{- i \o(t-t')} \times \nonumber \\ \times&\sum_{n=0}^{\infty} (n+\D-1/2)\tilde{G}_{\n+1/2}[z,|x_{\perp}'|] \frac{1}{|x_{\perp}'|^{\D}} \frac{P^{\D-1}_{n+\D-1}(\tanh x)}{\cosh x},
\end{align}
In appendix \ref{bulktodprop}, we show that this reduces to the well-known pure-AdS bulk-to-boundary propagator when the defect is switched off (in which case $\kappa = 1$ and $\n = n+\D-1$).

The correlation functions in (defect) CFTs are typically presented in position space. So we would like to perform the inverse Fourier
transform of \eqref{bulkboundprop}. To this end, we note the Gegenbauer's formula \cite{watson1995treatise} 
\begin{align}
&\frac{H_{\n+m}^{(1)}(\o z)}{(\o z)^{\n}} \frac{J_{\n+m}(\o {x'_{\perp}})}{(\o x'_{\perp})^{\n}} \! = \!
\nonumber\\
& \qquad \qquad
\frac{2^{\n-1} m! \G(\n)}{\pi \G(2 \n +m)}
\int_0^{\pi} \!\!  \frac{H_{\n+m}^{(1)}(|\o|\sqrt{z^2 + {x^{'2}_{\perp}} - 2 z x'_{\perp}\cos {\theta}})}{\o^{\n}(z^2 + x^{'2}_{\perp}
- 2 z x'_{\perp}\cos {\theta})^{\n/2}} C_m^{\n}(\cos \theta) \sin^{2\n}\! \theta d \theta,
\end{align}
where $C_m^{\n}(\cos \theta)$ denotes the Gegenbauer function.
For the computation in Euclidean signature the same formula holds if we exchange $J_{\n+m}$ by $I_{\n+m}$ and $H_{\n+m}$ by $K_{\n+m}$
on the both sides of the equation.
Using this relation (with $m=0$) in \eqref{bulkboundprop} and Fourier transforming back, we are left with an integral over $\theta$ which can be expressed in
terms of the hypergeometric function. Using \cite{bateman1953higher}
\be
{}_2F_1\Big(\n+1,\n+1,2\n+2,\frac{2}{1-z}\Big) = 2^{\n+1} \frac{\G[\n+3/2]}{\sqrt{\pi}\G[\n+1]} (z-1)^{\n+1} Q_{\n}(z),
\ee
we are left with a sum
\begin{align}
K_{\D}= &\Big(-\frac{1}{2} \Big)^{\D-1} \frac{1}{\pi \kappa^\Delta \Gamma(\D-1)} \frac{1}{|x_{\perp}'|^{\D}} \nonumber\\
&\qquad \sum_{n=0}^{\infty}(n+\D-1/2)
Q_{\n}\Big(\frac{z^2 + x^{'2}_{\perp} +(t-t')^2}{2 z x'_{\perp}}\Big) \frac{P^{\D-1}_{n+\D-1}(\tanh x)}{\cosh x}.
\end{align}
For the undeformed case, $\n = n+\D-1$, and we use the identity (\cite{magnus1966formulas}, see also \cite{2011arXiv1107.2680S})
\be
(z-\cos (\phi))^{-k-1/2} = \frac{2^{k+1/2}}{\pi^{1/2}} \frac{\Gamma(k)}{\Gamma(k+1/2)}\sum^{\infty}_{n=0} (k+n) C_{n}^k (\cos (\phi)) Q_{n+k-1/2}(z),
\quad \text{Re} \ k >-\frac{1}{2}
\ee
to check that the position-space propagator \eqref{positionspaceprop} is indeed reproduced.
For the deformed case, the sum cannot be evaluated in closed form. As we will see later, each term in the sum corresponds
to the expected contribution of a descendant of the dual operator $\mathcal{O}$.

\subsection{Correlation functions}

\begin{figure}[t]
\includegraphics[width=0.48\textwidth]{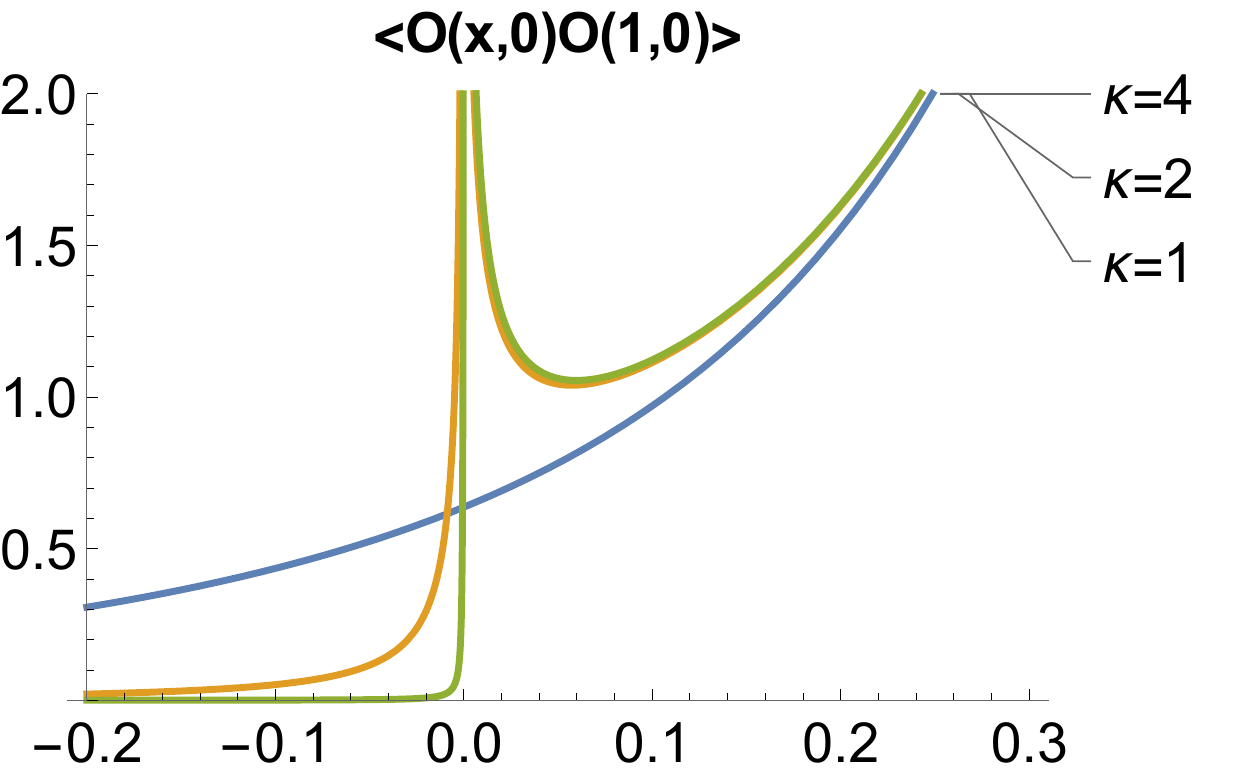} \ \ \ \ \includegraphics[width=0.4\textwidth]{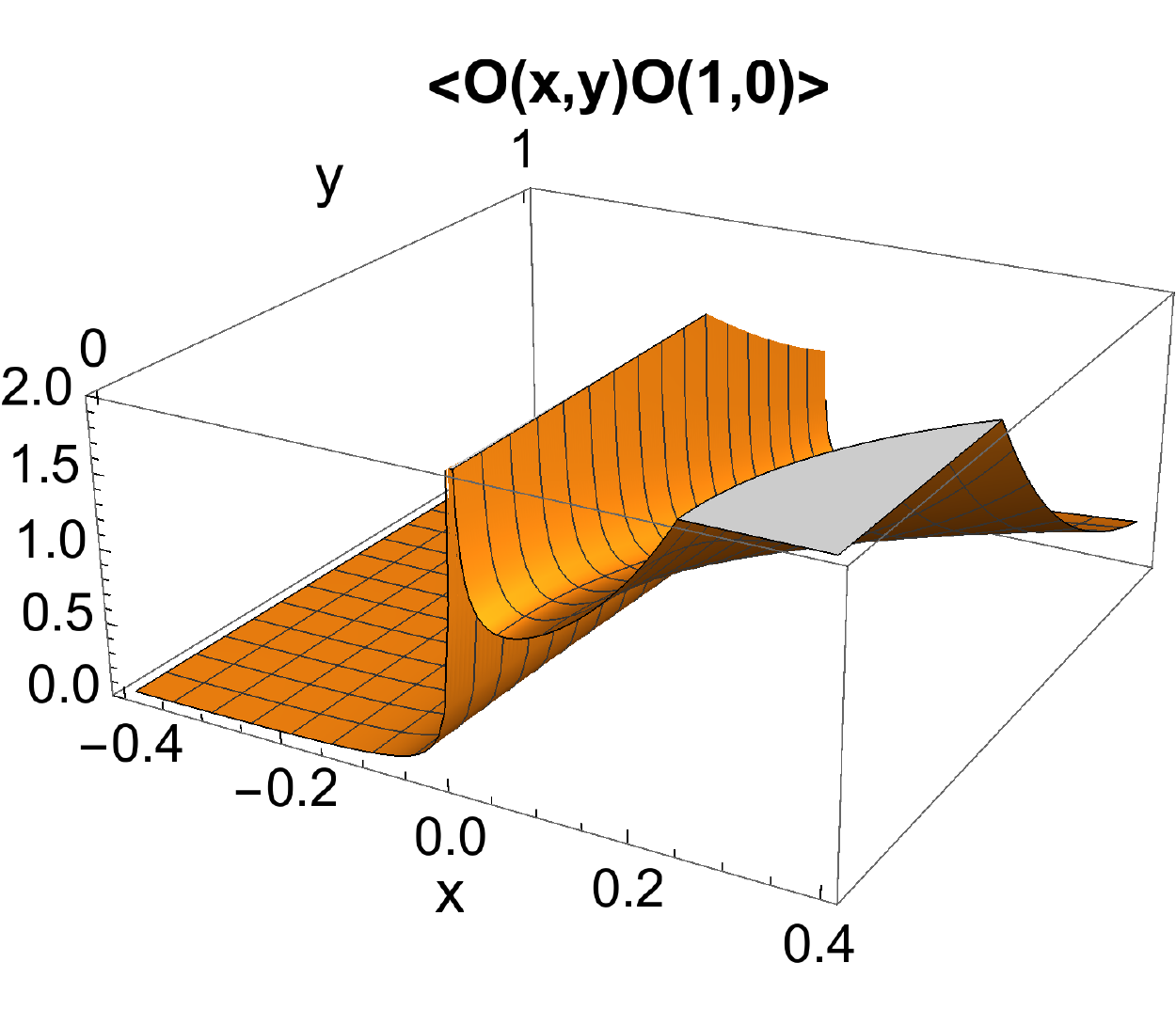}
\small \caption{ \small
Plot of $\< \mathcal{O}(x,0) \mathcal{O}(1,0) \> $ for $\kappa=1,2,4$ and $\Delta=2$ (left)
and plot of $\< \mathcal{O}(x,y) \mathcal{O}(1,0) \> $ for $\kappa=2$ and $\Delta=2$ (right). \label{plot}
}
\end{figure}

The bulk field $\Phi$ is dual to an operator $\mathcal{O}$ in the dual CFT. Solving the field equation on an
asymptotically locally-AdS  background near the boundary, one obtains the asymptotic solution
\be
\Phi = e^{(\D-d)r} \Big(\phi_{(0)} + \ldots +e^{(d-2 \D)r} \phi_{(2\D-d)} +\ldots \Big).
\ee
The expectation value $\< \mathcal{O} \>$ in the presence of the source $\phi_{(0)}$ is given by
\be
\< \mathcal{O} \>_{\phi_{(0)}} = (d-2\D) \phi_{(2\D-d)}[\phi_{(0)}] + \text{local (scheme-dependent) terms}.
\ee
At the same time, the bulk solution with prescribed source $\phi_{(0)}$ can be obtained with the help of the bulk-to-boundary propagator as
\be
\Phi(x) = \int d^d y K_{\D}(x,y) \phi_{(0)}(y).
\ee
Since we know the bulk-to-boundary propagator for the defect geometries, we can directly use the last formula
to extract the two-point function by extracting the corresponding term in the near-boundary expansion and differentiating the
one-point function with respect to the source. After performing these steps, we obtain the final result:
\begin{align}
\label{2ptcorr}
\< \mathcal{O}(x_{\perp},t) \mathcal{O}(x_{\perp}',t') \> =& \frac{2^{3-2\D}}{\pi \kappa^{2\Delta} [\G(\D-1)]^2} \frac{1}{|x_{\perp} x'_{\perp}|^{\D}} \times \nonumber \\ &\times \sum_{n=0}^{\infty} \sign(\xi)^n(n+\D-1/2)\frac{\G(n+2\D-1)}{\G(n+1)}Q_{\n}(|2\xi +1|),
\end{align}
where $\xi$ was defined in \eqref{defxi} and the above formula applies regardless of whether the
insertions are on the same side of the defect ($\xi>0$) or not ($\xi<0$). \eqref{2ptcorr}
has exactly the functional dependence  \eqref{boundaryblocks} obtained in \cite{McAvity:1995zd}.
We recognize each term in the sum as a contribution from a particular descendant $\hat{\mathcal{O}}_n=\partial^n_{x_{\perp}} \hat{\mathcal{O}}$.
Plots can be found in figure \ref{plot}. We see in particular that the singularity when the two insertions are brought close to each other
is not sensitive to the deformation parameter. In the presence of the defect, the correlator is also singular as one of the
operators approaches the defect. Finally, there is a ``screening" effect when the operators are placed on opposite sides of the defect, i.e. the correlator decays to zero quickly behind the defect.

Moreover, now we can read off the BOPE coefficient as
\be
b_{\mathcal{O} \hat{\mathcal{O}}_n}^2 = 2^{2(\D-\hat{\D}_n)} \frac{\Gamma(\hat{\D}_n)}{\Gamma(\hat{\D}_n+1/2)} \frac{2^{3-2\D}}{\sqrt{\pi} [\G(\D-1)]^2} \frac{1}{\k^{2\D}} (n+\D-1/2)\frac{\G(n+2\D-1)}{\G(n+1)}.
\ee

The formula \eqref{2ptcorr} is the main result of this section. We already know
that it does reduce to the usual CFT result when the defect goes away. Moreover the final result appears in a form which makes the BOPE block
decomposition explicit. There is one more check we can perform. Let's consider the case in which the operator insertions are situated far away from
the defect but close to each other. This corresponds to the limit $\xi \rightarrow 0$. In this limit we expect to find the decomposition of the two-point
correlation functions in terms of the usual OPE, i.e. we should find the singularities of the form $1/\xi^{\Delta}$. Notice that each term in \eqref{2ptcorr}
diverges only logarithmically as $\xi \rightarrow 0$. Thus, to obtain the scaling behavior, we need to resum infinitely many terms in \eqref{2ptcorr}.
To achieve this, we can use asymptotic expressions for large $n$.

At large $n$ ($n \gg \k$) we can approximate $\n \sim n/\k$. The asymptotic expansion of the Legendre function at large $n$ is
directly obtainable from the hypergeometric series representation
\be
Q_{\n}(z) = \sqrt{\frac{\pi}{2}} \frac{\G[\n+1]}{\G[\n+3/2]} \frac{[z - (z^2-1)^{1/2}]^{\n+1/2}}{(z^2-1)^{1/4}} {}_2F_1\Big(\frac{1}{2}, \frac{1}{2};\n+\frac{3}{2}; \frac{\sqrt{z^2-1}-z}{2\sqrt{z^2-1}}\Big).
\ee
Using this and Stirling's approximation for $\G$-functions we can perform a sum in \eqref{2ptcorr} from some large $N$ to infinity.
In the following, we will ignore numerical prefactors (except for the dependence on the deformation parameter $\k$) since we are after scaling behavior only. The computation goes as follows:
\begin{align}
&\sum_{n=N}^{\infty}(n+\D-1/2)\frac{\G(n+2\D-1)}{\G(n+1)}Q_{\n}\Big(2\xi +1\Big) \nonumber \\
& \qquad \qquad \qquad   \sim \k^{1/2}\Big(\frac{1}{\xi^{1/4}}+ \mathcal{O}(\xi^{1/4}) \Big)\sum_{n=N}^{\infty} n^{2 \D - 5/2} \Big(1+{\mathcal{O}}\Big(\frac{1}{n}\Big)\Big) \zeta^{n/\k} \nonumber \\ &
\qquad \qquad  \qquad \qquad  \times \Big( 1-\frac{\k}{4 n} \frac{\zeta}{4 \sqrt{\xi(\xi+1)}} + \mathcal{O}\Big(\frac{1}{n^2}\Big) \Big),
\end{align}
where we defined $\zeta = 2 \xi + 1 - 2\sqrt{\xi(\xi+1)} \sim 1 - 2 \sqrt{\xi}$. The leading contribution can be resummed to give\footnote{Alternatively one could approximate the sum by an integral, which in turn can be evaluated in terms of the exponential integral function. However the sum under consideration does reduce to a well-studied special function.}
\be
\label{leadingsum}
\frac{\k^{1/2}}{\xi^{1/4}}\sum_{n=N}^{\infty} n^{2 \D - 5/2}  \zeta^{n/\k} = \frac{\k^{1/2}}{\xi^{1/4}} \zeta^{N/\k} \Phi[\zeta^{1/\k},\frac{3}{2}-2\D,N],
\ee
where
\be
\Phi[z,s,v]=\sum_{n=0}^{\infty} \frac{z^n}{(v+n)^s}
\ee
is the so-called Lerch transcendent \cite{bateman1953higher}. The asymptotic behavior of the Lerch transcendent as $\zeta \rightarrow 1$ can be deduced from \cite{bateman1953higher}
\be
\Phi[z,s,v] \sim \frac{\G[1-s]}{(1-z)^{1-s}}.
\ee
Plugging this into the \eqref{leadingsum} gives
\be
\frac{\k^{1/2}}{\xi^{1/4}} \frac{1}{(1-\zeta^{1/\k})^{2\D-1/2}} \sim \frac{\k^{2 \D}}{2^{2\D-1/2}}\frac{1}{\xi^{\D}},
\ee
which is exactly the identity contribution to the OPE in the bulk CFT!

It is easy to see from \eqref{2ptcorr} that there are no more contributions at the leading order.  In principle, one can now proceed
systematically and discover additional contributions with subleading scaling behavior.  These should be there since
the one-point functions are in general non-zero in the presence of the defect. However, this procedure
becomes prohibitively difficult even at the next order ($1/\xi^{\D-1/2}$). One reason  is that in the expansion of the hypergeometric function
there are infinitely-many terms contributing at each order.

\section{Discussion}

In the present paper, we have studied a particular family of top-down supersymmetric Janus solutions which asymptote to $AdS_3\times S^3$ and
are dual to two-dimensional defect conformal field theories. We have identified a scalar perturbation in the bulk which decouples at the leading order
and solved the linearized field equation.
Taking advantage of the simple structure of the background solution,
we managed to construct explicitly the bulk-to-bulk propagator and to extract the two-point correlation function.
The functional form of this correlator is not fixed by the symmetry alone. Rather, it involves an arbitrary
function of an invariant cross-ratio. The final expression for the two-point correlation function is represented as a sum over boundary
conformal blocks and makes manifest the data present in the OPE of the operator with the defect (dimensions and OPE coefficients). Finally,
we have partially checked that our final result is crossing symmetric.

There are several directions one can now pursue.
The theory which is holographically dual to the Janus solution under consideration is known in detail.
Moreover, we have identified the microscopic description of the dual operator in the dual field theory. Hence,
it is in principle possible  to compute the same correlation function  in the field theory at weak coupling.

Our analysis of the linearized fluctuations around the supergravity solution was not completely systematic.
In particular, there is certainly room for other fluctuations to decouple at the leading order.
One possibility would be to consider fluctuations of the B-field with the indices along the four-torus.
Additionally, it would be natural to consider perturbations around the background solution that involve more than one
field. Indeed, treating perturbations of this sort is required for computing holographically correlators involving the energy-momentum
tensor or other conserved currents.

Furthermore, it would be natural to attempt extending our results
to Janus solutions in higher dimension and with more asymptotic regions. Perturbations around the supersymmetric
$AdS_5 \times S^5$ Janus solutions have been studied in \cite{Bachas:2011xa}, where particular modes that obey
separable differential equations were identified. However, these equations are not of the hypergeometric type (they are known in the literature
as Heun's equations) and  in general their solutions can be studied  only numerically. In case of solutions with more
than two asymptotic regions, the differential equations would not be separable even in the
asymptotically $AdS_3 \times S^3$ case.

Finally, further investigating finite-temperature solutions would be a particularly promising direction.
Since we  managed to solve for the linearized modes at finite temperature,
one can proceed to compute quasi-normal modes and/or correlation functions, which are hard to compute even at weak coupling.
It is not clear to us if this can be done as explicitly as for the zero-temperature case, but it is certainly possible to find solutions at least numerically.
However, in order to proceed in this direction, a better understanding of the global properties of the finite-temperature solutions is required.

\section*{Acknowledgments}
We would like to thank Marco Meineri and Michael Gutperle for useful discussions and Andy O'Bannon for helpful comments on the draft.
The work of MC is supported in part by the Knut and Alice Wallenberg Foundation under grant KAW 2013.0235r.

\appendix

\section{Bulk-to-boundary propagator in pure anti-de Sitter}
\label{bulktodprop}

It might be useful to demonstrate how the well-known bulk-to-boundary propagator in pure $AdS$ can be represented using our defect modes. To do that we start with the expression
\begin{align}
\phi(x_0,\vec{x}) = \int d^d\vec{x}' K_{\D}(x_0, \vec{x}, \vec{x}') \phi_0 (\vec{x}'),
\end{align}
where the bulk-to-boundary propagator in Poincar\'e coordinates is\footnote{Recall that the propagator is normalized in such a way that
\be
\lim_{x_0 \rightarrow 0} x_0^{\D-2} K_{\D}(x_0, \vec{x}, \vec{x}') = \delta(\vec{x} - \vec{x}').
\ee}
\be
\label{positionspaceprop}
K_{\D}(x_0, \vec{x}, \vec{x}') = C \frac{x_0^{\D}}{(x_0^2 + (\vec{x}-\vec{x}')^2 )^{\D}}, \qquad C = \pi^{-d/2}\frac{\G(\D)}{\G(\D-d/2)}.
\ee
Here $\phi_0 (\vec{x}')$ represents the source on the boundary of AdS. Next we decompose boundary directions as $\vec{x}=(x_{\perp},\vec{y})$ and similarly for the location of the source $\vec{x}'$ and go to hyperbolic slicing by defining new coordinates $z$ and $x$ by
\be
x_0 = \frac{z}{\cosh x}, \qquad x_{\perp} = z \tanh x.
\ee

Now we Fourier-transform along the directions of the defect to obtain (in Euclidean signature)\footnote{To evaluate this multi-dimensional Fourier transform we note that for a spherically symmetric function $f(\vec{y}) = f(|\vec{y}|)$ the Fourier transform is related to the so-called Hankel transform through
\be
\int d^n k e^{- i \vec{k} \cdot \vec{y}} f(\vec{y}) = (2 \pi)^{n/2} k^{1-n/2} \int_0^{\infty} r^{n/2} f(r) J_{\frac{n}{2}-1}(kr)dr.
\ee}
\be
K_{\D}(x_0, \vec{x}, \vec{x}')= \frac{2}{\sqrt{\pi} \Gamma(\D-d/2)}\frac{z^{\D}}{\cosh^{\D}{x}} \frac{1}{(2 \pi)^{d-1}}\int d^{d-1}k\, e^{-i \vec{k} \cdot (\vec{y}-\vec{y}')} \Big(\frac{ k^2}{2}\Big)^{\D+\frac{1-d}{2}}  \frac{K_{\D+\frac{1-d}{2}}(k w)}{(k \o)^{\D+\frac{1-d}{2}}},
\ee
where
\be
w = \sqrt{z^2 - 2 z \tanh xx_\perp' + x_\perp'^2},
\ee
and $K_{\n}$ denotes the modified Bessel function.
Now we use the Gegenbauer's addition theorem for Bessel functions which says \cite{watson1995treatise}
\be
\frac{K_{\n}(k w)}{(k w)^{\n}} = 2^{\n} \Gamma(\n)\sum_{n=0}^{\infty}(\n+n) \frac{K_{\n+n}(k z)}{(k z)^{\n}} \frac{I_{\n+n}(k x_\perp')}{(k x_\perp')^{\n}} C_n^{\D-1/2}(\tanh x)
\ee
for $z>|x_\perp'|$. Whereas for $z<|x_\perp'|$ the arguments of the Bessel functions should be exchanged.
In Lorentzian signature we would need \cite{watson1995treatise} 
\be
\frac{H_{\n}(k w)}{(k w)^{\n}} = 2^{\n} \Gamma(\n)\sum_{n=0}^{\infty}(\n+n) \frac{H_{\n+n}(k z)}{(k z)^{\n}} \frac{J_{\n+n}(k x_\perp')}{(k x_\perp')^{\n}} C_n^{\D-1/2}(\tanh x).
\ee
Thus the bulk-to-boundary propagator in pure AdS can be expressed as
\begin{align}
K_{\D}(x_0, \vec{x}, \vec{x}') = \frac{2}{\sqrt{\pi}} &\frac{\Gamma(\D-d/2+1/2)}{\Gamma(\D-d/2)}\frac{z^{\D}}{\cosh^{\D}{x}}
\frac{1}{(2 \pi)^{d-1}}\int d^{d-1}k\, e^{-i \vec{k} \cdot (\vec{y}-\vec{y}')} k^{2\D+1-d}  \times\nonumber \\ \sum_{n=0}^{\infty}(\D+\frac{1-d}{2}+n)
& \left[  \theta(z-x_\perp')\frac{K_{\D+\frac{1-d}{2}+n}(k z)}{(k z)^{\D+\frac{1-d}{2}}} \frac{I_{\D+\frac{1-d}{2}+n}(k x_\perp')}{(k x_\perp')^{\D+\frac{1-d}{2}}}
+\nonumber \right.  \\ +  & \left. \theta(x_\perp'-z) \frac{I_{\D+\frac{1-d}{2}+n}(k z)}{(k z)^{\D+\frac{1-d}{2}}} \frac{K_{\D+\frac{1-d}{2}+n}(k x_\perp')}{(k x_\perp')^{\D+
\frac{1-d}{2}}}  \right]C_n^{\D+\frac{1-d}{2}}(\tanh x).\nonumber
\end{align}
Finally notice that for integer $\D-d/2$ the Gegenbauer function is related to the associated Legendre functions by \cite{bateman1953higher} \begin{align}
C_n^{\D-1/2}(\tanh x) 
= \frac{\G(\D)}{\G(2\D-1)} (-2 \cosh x)^{\D-1} P^{\D-1}_{n+\D-1} (\tanh x).
\end{align}
(For non-integer $\D$ there is also $Q^{\D-1}_{n+\D-1} (\tanh x)$ appearing in the final equality). Thus we see that the bulk-to-boundary propagator can be expressed using the defect modes derived before.

Let us now specialise to $d=2$ and integer $\D$, so that $\vec{y}$ becomes just $t$ and $\vec{k}$ reduces to $\omega$. We get
\begin{align}
\label{adsbulkboundprop}
K_{\D}(x_0, \vec{x}, \vec{x}') = &\Big( -\frac{1}{2} \Big)^{\D-1} \frac{1}{\Gamma(\D-1)}\frac{1}{\pi}\int_{-\infty}^{\infty} d\o e^{- i \o (t-t')}  \times \\ \times&\sum_{n=0}^{\infty}(n + \D - 1/2) \tilde{G}^{Eucl}_{n+\D-1/2}[z,x_\perp'] \frac{1}{x_\perp'^{\D}}\frac{P^{\D-1}_{n+\D-1}(\tanh x)}{\cosh x},\nonumber
\end{align}
where $\tilde{G}^{Eucl}_{n+\D-1/2}[z,x_\perp']$ is obtained from \eqref{zproblem} by analytically continuing $\omega^2 \rightarrow -\omega^2$ which effectively replaces Bessel functions according to $J_{\n} \rightarrow I_{\n}$ and $H_{\n} \rightarrow K_{\n}$ ($K_{\n}(|\o|z)$ is regular at large $z$). \eqref{adsbulkboundprop} indeed coincides with the pure AdS limit of \eqref{bulkboundprop}.

\newpage

\bibliographystyle{JHEP}
\bibliography{literature}

\end{document}